\documentclass[aps,prl,
preprint,
amsmath]{revtex4}
\begin{document}
\title{Recombination and Lifetimes of Charge Carriers in Semiconductors}
\author{I.N.~Volovichev}
\affiliation{Institute for Radiophysics and
Electronics, National Academy of Science of Ukraine, Kharkov, 61085,
Ukraine.}
\author{G.N.~Logvinov}
\affiliation{SEPI, ESIME Culhuacan,
Instituto Politecnico Nacional, Av.~Santa Ana 1000, Col.~San Francisco,
Culhuacan, C.D. 04430, D.F., M\'{e}xico.}
\author{O.Yu.~Titov}
\author{Yu.G.~Gurevich}
\email{gurevich@fis.cinvestav.mx}
\affiliation{Depto.~de F\'{\i}sica, CINVESTAV---IPN, Apdo.~ Postal
14--740, 07300 M\'{e}xico, D.F., M\'{e}xico}
\begin{abstract}

In this Communication, it is shown that models used for describing
generation and recombination of electrons and holes lead to
disagreements with Maxwell's electrodynamics. Self-consistent
expressions, more adequately depicting the actual physical
processes of electron-hole recombination in semiconductors are
obtained. It is shown that the electron and hole lifetimes can be
defined correctly only for the special cases when the electron and
hole nonequilibrium concentrations are the same, these lifetimes
being equal. The influence of temperature inhomogeneity on the
recombination is also considered. The recombination rate for hot
electrons is obtained in the case when the electron and hole
temperatures differ.
\end{abstract}

\maketitle

Generation and recombination phenomena involving electron and
holes are very important processes in any nonequilibrium
thermodynamics, and particularly for the study of electron
transport of any kind, such as the charge, heat and particle
transport. Processes like those are also essential for the
designing of any solid state device.

Nevertheless, all models of generation and recombination processes
used today contradict fundamental physics such as those described
by Maxwell's equations.

Most of the electron and hole transport phenomena usually are
accompanied by appearance of nonequilibrium carriers of charge
(see Ref.~\cite{0,1,2,3,4,5,6,7}).

Their motion is described by the set of continuity equations for the
electron $\mathbf{j}_{n}$, and hole $\mathbf{j}_{p}$ current densities
(Ref.~\cite{3,7}),
\begin{equation}
\begin{split}
\frac{\partial n}{\partial t} & =
g_{n} + \frac{1}{e}\operatorname{div} \mathbf{j}_{n} - R_{n} ,\\
\frac{\partial p}{\partial t} & =
g_{p} - \frac{1}{e}\operatorname{div} \mathbf{j}_{p} - R_{p} ,
\end{split}
\end{equation}
and by the Poisson equation (see Ref.~\cite{5,7}),
\begin{equation}
\operatorname{div} \mathbf{E} = \frac{4\pi}{\varepsilon}\rho.
\end{equation}
Here $n$ and $p$ are the total electron and hole concentrations,
$g_{n}$ and $g_{p}$ are the electron and hole external generation
rates, $\mathbf{E}$ is the electric field, $\rho$ is the bulk
electric charge density, $e$ is the hole charge, $\varepsilon$ is
the permittivity, $R_{n}$ and $R_{p}$  are the electron and hole
recombination rates.

Usually, $R_{n}$ and $R_{p}$ are specified by the following
equations (see Refs.~\cite{0,1,2,3,5,7,8}):
\begin{equation}
R_{n} = \frac{\delta n}{\tau_{n}}, \qquad
R_{p} = \frac{\delta p}{\tau_{p}},
\end{equation}
where $\delta n = n - n_{0}$, $\delta p = p - p_{0}$, $n_{0}$ and
$p_{0}$ are the equilibrium electron and hole concentrations;
$\tau_{n}$ and $\tau_{p}$ are the electron and hole lifetimes.

Eqs.~(3) are obtained for small deviations of the total concentrations
from the equilibrium ones, $\delta n,\delta p \ll n_{0},p_{0}$.

Subtracting the second equation from the first one [see Eqs.(1)], we
obtain that
\begin{equation}
e(g_{n} - g_{p}) + \operatorname{div}(\mathbf{j}_{n} +
\mathbf{j}_{p}) - e(R_{n} - R_{p}) = e \frac{\partial}{\partial t}\left(n-p\right).
\end{equation}

At the same time, it is easy to obtain the continuity equation for the total current
$\mathbf{j} = \mathbf{j}_{n} + \mathbf{j}_{p}$ from the Maxwell's
equations,
\begin{equation}
\operatorname{div} \mathbf{j} = e \frac{\partial}{\partial t}\left(n-p+n_t\right),
\end{equation}
where $n_t$ is the concentration of captured electrons.

Comparing  Eqs.~(4) and (5)  we get that
\begin{equation}
g_{n} - g_{p} = R_{n} - R_{p} -\frac{\partial n_t}{\partial t}.
\end{equation}

After substituting Eq.~(3) into Eq.~(6) we will obtain an
unphysical result. In this case we have one more correlation
between $\delta n$ and $\delta p$ that overdetermines the set of
Eqs.~(1)-(2). The latter proves the incorrectness of Eqs.~(3). It
is interesting to note, as it follows from Eq.~(6), in general
case the rates $R_n$ and $R_p$ as functions of $n$ and $p$ are
dependent on the external generation rates $g_n$ and $g_p$.

At the same time, starting from the well known equation for the band-band
recombination (see Ref.~\cite{1}),
\begin{equation}
R = \alpha(np - n^{2}_{i}) ,
\end{equation}
it is easy to obtain that Eq.~(6) will be satisfied identically.
In Eq.~(7)  $n^{2}_{i} = n_{0} p_{0}$, $n_{i}$ is the electron
concentration in intrinsic semiconductor at the equilibrium
temperature $T_{0}$ and $\alpha$ is the capture coefficient.

Actually, $g_{n} = g_{p}$, and $R_{n} = R_{p}$ if the impurity
levels (traps) do not take part in the generation and
recombination processes ($n_t= \rm const$). From the above,
instead Eq.~(3), the equation for the band-band recombination must
be
\begin{equation}
R_{n} = R_{p} = R = \frac{\delta n}{\tau_{n}}+ \frac{\delta
p}{\tau_{p}}. \end{equation}

Here $\tau_{n} = (\alpha p_{0})^{-1}$, and $\tau_{p} = (\alpha
n_{0})^{-1}$. As usually, we have assumed that $\delta n, \delta p \ll
n_{0},p_{0}$.

It is necessary to emphasize that even if parameters $\tau_{n}$ and
$\tau_{p}$ have dimension of time they are not the lifetimes of
the nonequilibrium carriers. That follows straightforwardly from Eq.~(1)
after taking into account Eq.~(8).

The lifetime of the charge carriers can be introduced only if $\delta n =
\delta p$. The latter equality takes place if and only if all
characteristic lengths of physical processes are much large than the Debye
radius (the quasineutrality condition, see Ref.~\cite{5}). Moreover, the
Maxwell's time of the carriers have to be much less than the transition
times and other characteristic times (see Ref.~\cite{5,9}).

If $\delta n = \delta p$, then the lifetime for electrons as well
as for holes will be given by one unique expression:
$\tau=\tau_{n}\tau_{p}/(\tau_{n}+\tau_{p})$.

Let us note that even in the case of intrinsic semiconductors ($n_{0} =
p_{0}$, $\tau_{n}=\tau_{p}$), the band-band recombination does not provide
that $\delta n = \delta p$, contrary to widely spread opinion (see for
instance Refs.~\cite{1,2,5,7,8,9,10}. Actually, the presence of terms
$\operatorname{div}\mathbf{j}_{n}$ and $\operatorname{div}\mathbf{j}_{p}$
in Eq.~(1), in general makes impossible the equality $\delta n = \delta
p$.

As a rule, the electric current flowing through the semiconductor
structure heats this structure. In this case the temperature of a
sample becomes a function of the coordinates $T(\mathbf{r})$, see
Refs.~\cite{1,7,9}. In some cases (thermoelectric phenomena, and
devices based on these effects) the inhomogeneous temperature is
created willingly, with the help of external heaters and coolers
(Ref.~\cite{6}), or by heating with light (photothermal phenomena,
see Ref.~\cite{11}).

Not only in this case but more generally the electron $T_{n}$,
hole $T_{p}$ and phonon $T_{ph}$ temperatures are different
[$T_{n}(\mathbf{r}) \ne T_{p}(\mathbf{r}) \ne
T_{ph}(\mathbf{r})$], see Refs.~\cite{4,7}. For simplicity, let us
restrict ourselves to the case where the hole and the phonon
temperatures are equal: [$T(\mathbf{r}) = T_{p}(\mathbf{r}) =
T_{ph}(\mathbf{r})$]. At the same time we assume that the electron
temperature $T_{n}(\mathbf{r})$ differs from the phonon
temperature $T(\mathbf{r})$. Now, as it was shown in
Refs.~\cite{12,13}, we can obtain that
\begin{equation}
R = \alpha [T_{n}(\mathbf{r}),T(\mathbf{r})]np -
\alpha[T(\mathbf{r}),T(\mathbf{r})]n^{2}_{i} .
\end{equation}

The first term in Eq.~(9) corresponds to the  capture of the charge
carriers, and it depends on both temperatures. The second term
corresponding to the thermal generation depends only on  the phonon
temperature $T(\mathbf{r})$.

Let us now suppose the temperatures $T_n(\mathbf{r})$ and $T(\mathbf{r})$
are to be small differed from the equilibrium temperature $T_{0}$, i.e.,
\[
T_{n}(\mathbf{r}) = T_{0} + \delta T_{n}(\mathbf{r}), \text{  }
T(\mathbf{r}) = T_{0} + \delta T(\mathbf{r}), \text{  }\delta
T_{n},\text{  }\delta T  \ll T_{0}.
\]

Under  this assumption we can get  from Eq.~(9) that the recombination
rate
\begin{equation}
R = \frac{1}{\tau}
\left[
\frac{p_{0}}{n_{0} + p_{0}}\delta n + \frac{n_{0}}{n_{0} + p_{0}}\delta p
- \beta\delta T + \gamma(\delta T_{n} - \delta T)
\right] .
\end{equation}

Here $\beta = \frac{n_{0}p_{0}}{n_{0} +
p_{0}}\frac{1}{T_{0}}(3+\frac{\varepsilon_{g}}{T_{0}})$, $\gamma =
\left.\frac{1}{\alpha}\frac{\partial\alpha}{\partial
T_{n}}\right|_{T_{n} = T_{0}}\frac{n_{0}p_{0}}{n_{0} + p_{0}}$,
$\varepsilon_{g}$ is the energy gap. Let us note that in this
paper we use the energy system of units. From Eq.~(10) it is easy
to see that, even if $\delta n = \delta p$, the lifetime can not
be introduced.

Situation becomes even more complicated if recombination resulting
through the impurity centers (traps) is taken into account. Within
the framework of the Shockley-Read model and with the assumption
that the carriers of impurity centers are characterized by the
temperature $T(\mathbf{r})$, the recombination can be given by the
following equations, see Refs.~\cite{0,1,5,9,10}:
\begin{equation}
\begin{split}
R_{n} & = \alpha_{n}(T_{n},T)n(N_{t} - n_{t}) -
\alpha_{n}(T,T)n_{1}n_{t}, \\ R_{p} & = \alpha_{p}(T)pn_{t} -
\alpha_{p}(T)p_{1}(N_{t} - n_{t}).
\end{split}
\end{equation}

Here $N_{t}$ is the impurity concentration, $\alpha_{n}$  and
$\alpha_{p}$ are the electron and hole capture coefficients,
$n_{1} = \nu_{n}(T)e^{-\varepsilon_{t}/T}$, $p_{1} =
\nu_{p}(T)e^{\frac{\varepsilon_{t} -\varepsilon_{g}}{T}}$,
$\varepsilon_{t}$ is the impurity energy level, $\nu_{n}(T) =
\frac{1}{4}\left(\frac{2m_{n}T}{\pi\hbar^{2}}\right)^{\frac{3}{2}}$,
$\nu_{p}(T) =
\frac{1}{4}
\left(
\frac{2m_{p}T}{\pi\hbar^{2}}
\right)^{\frac{3}{2}}$, and
$m_{n}$ and $m_{p}$ are the electron and hole effective masses.
As it follows from Eqs.~(11), one more unknown value $n_{t}$ arises when
the recombination takes place through the impurity centers. Eq.~(6)
can be used just for its determination. Determining $n_{t}$
from Eq.~(6), and inserting it in Eq.~(11) we obtain the
recombination equation depending on $n$ and $p$.

Below, for simplicity we will assume that $g_{n} = g_{p}=0$. These
equalities take place if the nonequilibrium carriers are generated
due to injection, see. Refs.~\cite{1,2,3,10} or redistribution
(Ref.~\cite{6}) in semiconductor sample. One more example where
this condition is fulfilled is the surface generation of
photo-carriers in the sample, see Refs.~\cite{5,8,9}.

In this case Eq.~(6) serving  for determination $n_{t}$ for
steady-state ($\partial n_t/\partial t = 0$)  reduces to
\begin{equation}
R_{n} = R_{p} .
\end{equation}

Determining from this equation $n_{t}$ and inserting it in Eq.~(11) it is
easy to make sure that Eq.~(12) turns to identity.

Suggesting,  as it is made above, the deviation of all values to be
small and representing $n_{t}$ in the
form $n_{t} = n^{0}_{t} + \delta n_{t}$, where $n^{0}_{t}$ is the
equilibrium concentration of electrons located on the
impurity levels, we can obtain the recombination rate through the traps as
follows,
\begin{multline}
R_{n} = R_{p} = R = \frac{1}{\tau}
\left\{
\frac{N_{t} - n^{0}_{t}}{n_{0} + n^{0}_{1}}\delta n +
\frac{n^{0}_{t}}{p_{0} + p^{0}_{1}}\delta p
\right.\\
-
\frac{n^{0}_{t}}{n_{0} + n^{0}_{1}}\delta n_{1} -
\frac{N_{t} - n^{0}_{t}}{p_{0} + p^{0}_{1}}\delta p_{1}
\\
+
\frac{1}{\alpha_{n}}
\left.\frac{\partial\alpha_{n}}{\partial T_{n}}\right|_{T_{n}=T_{0}}
\frac{n_{0}(N_{t} - n^{0}_{t})}{p_{0} + p^{0}_{1}}
(\delta T_{e} - \delta T)
\\
+
\left[
\frac{1}{\alpha_{n}}
\left(
\left.
\frac{\partial\alpha_{n}}{\partial T_{n}}
\right|_{T_{n} = T_{0}} +
\left.
\frac{\partial\alpha_{n}}{\partial T}
\right|_{T = T_{0}}
\right)
\frac{n_{0}(N_{t} - n^{0}_{t}) - n^{0}_{1}n^{0}_{t}}{n_{0} + n^{0}_{1}}
\right.
\\
-
\left.\left.
\frac{1}{\alpha_{p}}
\left.
\frac{\partial\alpha_{p}}{\partial T}
\right|_{T = T_{0}}
\frac{p^{0}_{1}(N_{t} - n^{0}_{t}) - p_{0}n^{0}_{t}}{p_{0} + p^{0}_{1}}
\right]
\delta T
\right\}
\end{multline}

Here
\[
\frac{1}{\tau} = \frac{\alpha_{n}(T_{0},T_{0})\alpha_{p}(T_{0})(n_{0}
+ n^{0}_{1})(p_{0} + p^{0}_{1})}{\alpha_{n}(T_{0},T_{0})(n_{0} +
n^{0}_{1}) + \alpha_{p}(T_{0})(p_{0} + p^{0}_{1})}
.
\]

Obtaining Eq.~(13) we have taken into account that $n_{1}$ and $p_{1}$ can
be presented in the following form, $n_{1} = n^{0}_{1} + \delta n_{1}$,
$p_{1} = p^{0}_{1} + \delta p_{1}$. Here
$\delta n_{1} =
\frac{n^{0}_{1}}{T_{0}}\left(\frac{3}{2} +
\frac{\varepsilon_{t}}{T_{0}}\right)\delta T$
and
$\delta p_{1} =
\frac{p^{0}_{1}}{T_{0}}\left(\frac{3}{2} -
\frac{\varepsilon_{t} - \varepsilon_{g}}{T_{0}}\right)\delta T$.

Let us note that the concentrations of nonequilibrium
carriers $\delta n$ and $\delta p$ turn to zero at the strong
recombination ($\tau\to 0$) if only $\delta T = \delta T_{n} = 0$, as it
follows from Eqs.~(8), (10), and~(13). As a matter of fact, since $R$ is
always the finite value, the expression in braces (13) must vanish with
$\tau\to 0$. This implies that equality $\delta n = \delta p = 0$ takes
place only if $\delta T_{n} = \delta T = 0$, [see for example Eq.~(8)].

Summarizing, in this Communication we have shown that for the
depiction of generation and  recombination phenomena to be
consistent with Maxwell's equations, essential modifications are
required. According with our results, it is necessary to replace
the standard expressions for recombination involving lifetimes for
nonequilibrium electrons and holes with expressions obtained when
the definition of lifetime lost physical sense.

\section{Acknowledgements}

Authors thank C.~Terrero-Escalante for useful discussions.
This work has been partially supported by Consejo Nacional de Ciencia y
Tecnologia (CONACYT), Mexico.


\begin{thebibliography}{99}
\newpage
\baselineskip=30pt
\bibitem{0}
Peter T. Landsberg, {\em Recombination in Semiconductors}
(Cambridge University Press, 1991).
\bibitem{1}
D.A.~Neamen, {\em Semiconductor Physics and Devices: Basic Principles}
(IRWIN, Burr Ridge, Illinois, 1992).
\bibitem{2}
K.~Seeger, {\em Semiconductor Physics: An Introduction}
(Springer-Verlag, Berlin, 1991).
\bibitem{3}
J.~Sing, {\em Semiconductor Devices: Basic Principles}
(John Wiley and Sons, Inc., New York, 2001).
\bibitem{4}
{\em Hot Electron Transport in Semiconductors,} edited by L.~Reggiani
(Springer-Verlag, New York, 1985).
\bibitem{5}
I.~Auth, D.~Genzow, K.H.~Herrman,
{\em Photoelektrische Erscheinungen} (Akademie-Verlag, Berlin,1997).
\bibitem{6}
Yu.G.~Gurevich, O.Yu.~Titov, G.N.~Logvinov, O.I.~Lyubimov,
Phys.~Rev.~B {\bfseries 51}, 6999 (1995).
\bibitem{7}
E.~Scholl, {\em Nonequilibrium Phase Transitions in Semiconductors}
(Springer-Verlag, Berlin, 1987).
\bibitem{8}
D.K.~Schroder, {\em Semiconductor Material and Device Characterization}
(A Wiley-Interscience  Publication, New York, 1990).
\bibitem{9}
V.L.~Bonch-Bruevich, S.G.~Kalashnikov,
{\em Physik  fur Halbleiter}
(VEB Deutscher Verlagder Wissenschaften, Berlin, 1982).
\bibitem{10}
S.M.~Sze, {\em Physics of Semiconductor Devices} (Wiley, New York, 1981).
\bibitem{11}
H.~Vargas, L.~C.~M.~Miranda, Phys.~Rep. {\bfseries 161}, 43, (1988).
\bibitem{12}
E.M.~Conwell, {\em High Field Transport in Semiconductors}
(Academic Press, New York and London, 1967).
\bibitem{13}
Yu.G.~Gurevich, I.N.~Volovichev, Phys.~Rev.~B {\bfseries 60}, 7715
(1999).
\end{thebibliography}
\end{document}